\documentstyle[12pt,epsf]{article}
\begin{document}
\begin{titlepage}
\begin{flushright}
KUNS-1899\\
\end{flushright}

\begin{center}
\vspace*{10mm}

{\LARGE \bf
$D$-term Inflation and Nonperturbative \\
\vspace{1.5mm}
K\"ahler Potential of Dilaton
}
\vspace{12mm}

{\large
Tetsutaro~Higaki\footnote{E-mail address:
tetsu@gauge.scphys.kyoto-u.ac.jp},~
Tatsuo~Kobayashi\footnote{E-mail address:
kobayash@gauge.scphys.kyoto-u.ac.jp}
~and~Osamu~Seto\footnote{E-mail address: osamu@mail.nctu.edu.tw}
}
\vspace{6mm}

$^{1,2}${\it Department of Physics, Kyoto University,
Kyoto 606-8502, Japan}\\[1mm]

$^3${\it Institute of Physics, National Chiao Tung University,
Hsinchu, Taiwan 300, R.O.C.}

\vspace{6mm}

\vspace*{15mm}

\begin{abstract}
We study the $D$-term inflation scenario with
a nonperturbative K\"ahler potential of the dilaton field.
Although the FI term which leads an inflationary expansion is given by
the derivative of the K\"ahler potential with respect to the dilaton in
heterotic string models with anomalous $U(1)$,
the too large magnitude is problematic for a viable $D$-term inflation.
In this paper, we point out that the K\"ahler potential with a
nonperturbative term can reduce the magnitude of FI term to desired values
while both the dilaton stabilization and $D$-term domination in the
potential are
realized by nonperturbative superpotential.

\end{abstract}

\end{center}
\end{titlepage}

Within the framework of supergravity theory,
scalar fields, including an inflaton field, generally gain
the Hubble-induced masses during inflation,
if the potential energy is dominated by $F$-terms.
Violation of the slow-roll condition by these mass terms is known as
the $\eta$-problem. From this viewpoint,
the $D$-term inflation is an attractive scenario,
because the inflaton field does not gain such a large mass
if the $D$-term is dominant during inflation
\cite{Stewart:1994ts,inflation,D-inf}.
The Fayet-Iliopoulos (FI) term plays a role as the potential energy
in $D$-term inflation models.

The FI term is generated through anomalous $U(1)$
symmetries in four-dimensional string models.
Indeed, most of four-dimensional string models have
anomalous $U(1)$'s for both heterotic string
models \cite{DSW,Kobayashi:1996pb}
and type I string models \cite{typeI}.
These anomalies can be cancelled by the Green-Schwarz (GS)
mechanism \cite{Green:sg}, where certain
fields transform non-linearly under
anomalous $U(1)$ symmetries.
In heterotic string models, the gauge kinetic function
$f$ is obtained by the dilaton field S as
\begin{equation}
f=S,
\label{gauge-fn}
\end{equation}
up to Kac-Moody levels, that is,
the gauge coupling $g$ is determined by the vacuum
expectation value (VEV) of $S$ as $1/g^2 = \langle Re(S) \rangle$.
Under the anomalous $U(1)$ transformation,
the dilaton field $S$ transforms nonlinearly as
$S \rightarrow S + i \delta_{GS}\theta(x)$,
where $\theta(x)$ is a parameter of the anomalous $U(1)$
transformation and
the anomaly coefficient $\delta_{GS}$ is
obtained as \footnote{In type I models, twisted moduli fields
transform nonlinearly under anomalous $U(1)$ symmetries.}
\begin{equation}
\delta_{GS} = \frac{1}{192 \pi^2} {\rm Tr} (Q),
\end{equation}
with the summation of anomalous $U(1)$ charges, Tr $(Q)$.
That generates the dilaton-dependent FI term
\begin{equation}
\xi = \delta_{GS} K_S,
\label{FI-S}
\end{equation}
in the unit of $M_p = 1$, where $K_S$ is the first derivative
of the K\"alher potential $K$ with respect to $S$ and
$M_P$ denotes the reduced Planck scale.
Hereafter we use the unit $M_p =1$.

Thus, it seems possible to embed the $D$-term inflation scenario
in string models.
However, we have problems.
One of problems is due to the fact that
the FI term $\xi$ and anomalous $U(1)$ gauge coupling
$g$ depend on $S$.
The tree-level K\"ahler potential of $S$ is obtained as
\begin{equation}
K_0(S + \bar S) = - \ln (S + \bar S).
\label{tree-K}
\end{equation}
With this form of the K\"ahler potential, the $D$-term
scalar potential $V_D$ during inflation behaves
$V_D = g^2\xi^2/2 \sim (S + \bar S)^{-3}$.
Only with this term, the dilaton rolls down rapidly
the potential to infinity, $Re(S) \rightarrow \infty$, and
the potential energy goes to zero, $V_D \rightarrow 0$.
The slow roll condition or the $D$-term inflation can
not be realized.
Therefore, the first problem is how to stabilize
the dilaton field during inflation.
For example, the dilaton can be stabilized by adding
the $F$-term scalar potential due to dilaton-dependent
nonperturbative superpotential terms \cite{King:1998uv}.\footnote{
Similarly, in the $D$-term inflation of
type I models, one can stabilize twisted moduli,
which determine FI terms \cite{Kobayashi:2003rx}.}

The second problem is concerned about the magnitude of $\xi$.
The GS coefficient $\delta_{GS}$ is model-dependent, and
explicit models lead to $\delta_{GS} =
O(10^{-1}) - O(10^{-4})$ \cite{Kobayashi:1996pb}.
On the other hand, the magnitude of the anisotropy of
the cosmic microwave background (CMB) requires
$\xi^{1/2} \leq O(10^{15} - 10^{16})$ GeV \cite{D-inf}.
Thus, explicit heterotic string models seem to have much larger
value of the FI term $\xi$.
Until now, just a few models leading to effectively small $\xi$ have
been studied \cite{Espinosa:1998ks,Lesgourgues:1998kj},
this seems still an open question.
Another way out of this problem is to consider $D$-term inflation
in type I string models \cite{Kobayashi:2003rx,Halyo:1999bq}.

Another problem is the generation of cosmic strings
in the true vacuum after inflation.
Its string tension is estimated as the FI term $\xi$ in the true vacuum.
If $\xi^{1/2} = O(10^{15}- 10^{16})$ GeV in the true vacuum,
cosmic strings besides inflaton lead to the density perturbation,
which is not consistent with the CMB observation
\cite{D-inf,Jeannerot:1997is,Endo:2003fr}.

Here we will study these problems in models with
nonperturbative K\"ahler potential of the dilaton field,
in particular, the stabilization problem and how to
reduce the magnitude of the FI term.
In the true vacuum, the dilaton stabilization has been studied
in models with its nonperturbative K\"ahler potential
in Refs.~\cite{Banks:1994sg}-\cite{Higaki:2003jt}.
In particular, in Ref. \cite{Higaki:2003jt}
it has been shown that one can stabilize $S$ with nonperturbative
K\"ahler potential in the true vacuum
such that $Re(S) = O(1)$ and $K_S$ is suppressed.
That corresponds to a suppressed value of $\xi$.
Thus, here we investigate the possibility for
stabilizing $S$ such that $K_S$ and $\xi$ are suppressed
during inflation.
We will also comment on the cosmic strings.

First, let us review briefly on the $D$-term
inflation \cite{inflation,D-inf}.
We consider a simple model with $U(1)$ gauge symmetry
and the non-vanishing FI term $\xi$.
This model includes three matter fields, $X$ and
$\phi_\pm$, and $\phi_\pm$ have $U(1)$ charges $\pm 1$ and
$X$ has no $U(1)$ charge.
The $U(1)$ $D$-term is written as
\begin{equation}
D = \xi + |\phi_+|^2 - |\phi_-|^2 .
\end{equation}
Here we take the charge normalization such that $\xi >0$.
In addition, we assume the following superpotential,
\begin{equation}
W = \lambda X \phi_+ \phi_-.
\end{equation}
Then, the scalar potential is written as
\begin{eqnarray}
V &=& \sum_i |\partial_i W|^2 + \frac{g^2}{2}D^2, \\
&=& \lambda^2 |X|^2(|\phi_+|^2 + |\phi_-|^2)
+ \lambda^2|\phi_+ \phi_ -|^2
+ \frac{g^2}{2}(\xi + |\phi_+|^2 - |\phi_-|^2 )^2.
\end{eqnarray}

The true vacuum corresponds to
\begin{equation}
X = \phi_+ =0, \qquad |\phi_-|^2 = \xi .
\end{equation}
In order to analyze the minimum of $V$ for a value of $X$ fixed,
we define $X_c$,
\begin{equation}
X_c \equiv \frac{g}{\lambda}\sqrt{\xi}.
\end{equation}
For $|X| < X_c$, the minimum corresponds to
\begin{equation}
|\phi_-|^2 = \xi - \frac{\lambda^2}{g^2}|X|^2, \qquad \phi_+=0 .
\end{equation}
On the other hand, for $|X| > X_c$, the minimum corresponds to
\begin{equation}
\phi_\pm =0 .
\end{equation}
In the latter case, the potential energy is obtained as
\begin{equation}
V = \frac{g^2}{2}\xi^2 ,
\label{VD-inf}
\end{equation}
and drives the inflationary expansion of the Universe,
where the radial part of $X$ is identified with the inflaton.
At the tree level, the inflaton $X$ has the flat potential.
The supersymmetry (SUSY) is broken during inflation because of
the non-vanishing $D$-term and
the scalar components of $\phi_\pm$ have masses squared,
\begin{equation}
m^2_\pm = \lambda^2 |X|^2 \pm g^2 \xi .
\end{equation}
Here the second term is the SUSY breaking mass squared
by non-vanishing $D$-term, while the first term is
the supersymmetric mass squared, which fermionic partners also have.
These mass splitting generates the one-loop effective potential,
\begin{equation}
V_{\rm 1-loop} = \frac{g^2}{2}\xi^2 \left( 1 +
\frac{g^2}{16\pi^2}\ln \frac{\lambda^2 |X|^2}{\Lambda^2} \right) ,
\end{equation}
where $\Lambda$ is the renormalization scale.
Thus, the potential for $X$ is slightly lifted, and the inflaton
$X$ slowly rolls down the potential.
The inflation ends when $X$ reaches at $X_c$ or $X_f$,
\begin{equation}
|X_f|^2 \equiv \frac{g^2}{8 \pi^2} M^2_p,
\end{equation}
where the slow roll condition is violated.

Next, let us consider the $D$-term inflation scenario from
the viewpoint of 4D effective theory of string theory.
Here, we use the tree-level K\"ahler potential (\ref{tree-K}) and
the gauge kinetic function (\ref{gauge-fn}).
Since the FI term and the gauge coupling are replaced with
$S$-dependent FI term (\ref{FI-S}) and $g^2 = 2/(S + \bar S)$ respectively,
the scalar potential
during inflation (\ref{VD-inf}) is rewritten as
\begin{eqnarray}
V &=& \frac{(\delta_{GS}K_S)^2}{S + \bar S}
\label{VD-KS} \\
&=& \frac{\delta_{GS}^2}{(S + \bar S)^3} .
\end{eqnarray}
Now, two problems we mentioned above become clear.
The first one is that only with this potential
the dilaton rapidly runs away to infinity, $Re(S) \rightarrow \infty$,
and the inflation can not be realized.
In Ref.\cite{King:1998uv}, the dilaton stabilization
has been studied by adding the $F$-term scalar potential
generated by gaugino condensation as well as non-vanishing flux of the
antisymmetric tensor field $B_{\mu \nu}$.
The second problem is that the scale of the potential energy
is too high to produce the density perturbation with
an appropriate magnitude as long as $Re(S) = O(1)$ and
$\delta_{GS} = O(10^{-1}) - O(10^{-4})$.
These problems will be studied in our models with
nonperturbative K\"ahler potential.

In Ref.\cite{Higaki:2003jt}, the possibility for
suppressing $K_S$ has been studied by taking account of
nonperturbative effects on K\"ahler potential.
Actually, if we have additional terms in
the K\"ahler potential $K$ other than
the tree-level term $K_0$, the potential minimum of
Eq.(\ref{VD-KS}) can correspond to the point\footnote{See also 
Refs.~\cite{Arkani-Hamed:1998nu,Dine:1998qr}.} $K_S = 0$.
However, that implies vanishing FI term and that is not good,
because we can not realize the $D$-term inflation.
We need another contribution to lead to
a suppressed, but non-vanishing value of $K_S$.
Hence we will consider models with
nonperturbative corrections and study the
dilaton stabilization in order to lead to
a suppressed, but non-vanishing FI term.

The K\"ahler potential must also have the perturbative correction 
$K_p$.
We expect \cite{Arkani-Hamed:1998nu}
\begin{equation}
K_p(S + \bar S) = \frac{a}{S + \bar S} + \cdots,
\end{equation}
with $a=O(1/8\pi^2)$,
because of $\langle K_p/K_0 \rangle = O(g^2/16\pi^2)$.
We consider the case with $Re(S) =O(1)$.
Hence, it is expected that the perturbative correction 
would be irrelevant.
For the moment, we neglect $K_p$, but we will 
consider its effects after examining explicit models.

Nonperturbative effects in the K\"ahler potential
are still unknown.
Following Refs. \cite{Shenker:1990uf,Banks:1994sg},
we use the following Ansatz for the nonperturbative term,
\begin{equation}
K_{np}(S+\bar S ) = d(S + \bar S)^{p/2}e^{-b (S+\bar S)^{1/2}} ,
\end{equation}
where $p,b > 0$.
We also include the over-all moduli field $T$.
Then, the total K\"ahler potential is written as
\begin{equation}
K = K_0 (S+\bar{S}) + K_{np} (S+\bar{S}) -3 \ln (T + \bar T) + K_{X\bar
X}|X|^2
+ \cdots ,
\label{kahler}
\end{equation}
where the fourth term in the right hand side is the K\"ahler
potential of $X$.
Only with this K\"ahler potential, the scalar potential
during inflation is written as
$V = (\delta_{GS}K_S)^2/(S + \bar S)$.
Its minimum corresponds to $K_S =0$ other than
the runaway vacuum,
that is, we can not obtain a finite vacuum energy
deriving the inflation.
Hence, another contribution is necessary to realize
the dilaton stabilization leading to a suppressed, but
finite value of $K_S$.

As a such contribution, here we assume a nonperturbative
superpotential.
We consider two types of nonperturbative superpotentials.
One is written as
\begin{equation}
W_1 = B e^{-24\pi^2 S/B} + h .
\end{equation}
Here the first term is due to the gaugino condensation with
the 1-loop beta-function coefficient $B$, and
the second term is due to the non-vanishing flux.
The second type of superpotential is written as
\begin{equation}
W_2 = \alpha b_1 e^{-24\pi^2 S/b_1} + \beta b_2 e^{-24\pi^2 S/b_2} .
\end{equation}
This superpotential is generated by double gaugino condensations
with 1-loop beta-function coefficients $b_1,b_2$, that is,
the so-called racetrack type \cite{Krasnikov:jj}.
For simplicity, we have assumed that either $W_1$ or $W_2$
does not include $T$.
Then, the total scalar potential during inflation is written as
\begin{eqnarray}
V_T &=& V_F + V_D, \\
V_F &=& \frac{e^{K_{np}(s)+K_{X \bar X}|X|^2}}{s (T + \bar T)^3 K_{S
\bar S}}
|K_S W + W_S|^2 ,\\
V_D &=& \frac{(\delta_{GS} K_S)^2}{s}
\end{eqnarray}
where $s = (S + \bar S)$.

In order to realize the $D$-term inflation,
we have to stabilize $S$ such that $V_D \gg V_F$.
Note that $V_D$ does not include $T$.
Thus, if $S$ is stabilized satisfying $V_D \gg V_F$,
the potential for the moduli $T$ would be insignificant and
its dynamics also could be ignored.
Furthermore, we are interested in the model leading to
a suppressed value of $V_D$ compared with $\delta^2_{GS}/s^3$,
which is the vacuum energy in the model only with the tree-level 
K\"ahler potential.
Concerned about a stabilized value of $Re(S)$,
we are interested in models with $s = O(1)$,
but we do not restrict ourselves to some specific values,
e.g. $s= 4$.
The stabilization we are discussing happens in the false vacuum,
and in the true vacuum the dilaton potential and its stabilized value
would be different from those of false vacuum.

Here we show examples of the dilaton stabilization with
$V_D \gg V_F$. Note that the density perturbation generated by the inflaton
is same as that in a simple model by the condition, $V_D \gg V_F$.
For simplicity, we take $e^{K_{X\bar X}|X|^2}/(T + \bar T)^3 =1/2$
in the following analysis.

Fig.~1 shows the case using the superpotential $W_1$
with the parameters, $p=0$,$b=1$, $d=1$, $B=54$, $h= -1/40$ and
$\delta_{GS}=0.005$.
The almost flat light line corresponds to $V_D$,
and the dotted one corresponds to $V_F$.
The bold one corresponds to the total
potential $V_T$.
In this model, the dilaton is stabilized at
$s = 4.9$, where we have $V_D \gg V_F$.
Around this minimum, $V_F$ is very steep, while
$V_D$ is very flat.
Such situation plays a role in realizing $V_D \gg V_F$.
The total potential value is obtained as 
$V_T = 2.7 \times 10^{-7}$, and that is comparable to 
$\delta_{GS}^2/s^3$, which is the potential value 
derived only by the tree-level K\"ahler potential $K_0$.
In this case, the non-perturbative K\"ahler potential 
does not contribute to suppress the FI term.
\begin{figure}
\epsfxsize=0.7\textwidth
\centerline{\epsfbox{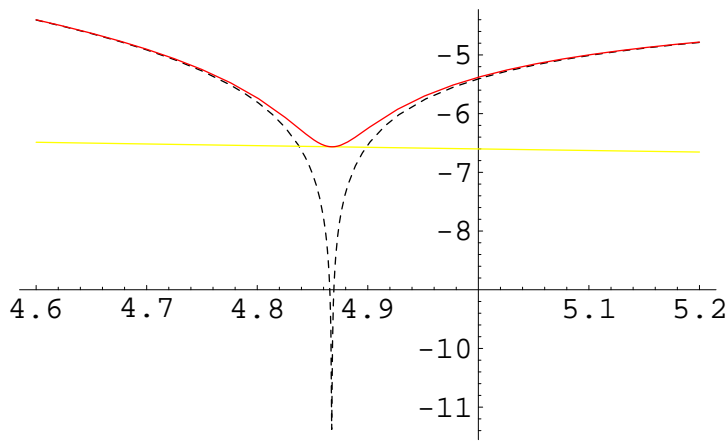}}
\caption{The potentials
for $W_1$ with parameters, $p=0,b=1,d=1,B=54,h=-\frac{1}{40}$ and
$\delta_{GS}=0.005$.
A horizontal axis represents $s$ and a vertical axis represents
potentials.
When $s\simeq 4.9$, $V_T \simeq 2.7 \times 10^{-7}$.
(light - $V_D$, dotted - $V_F$,
bold - $V_T$)}
\end{figure}

In order to obtain more suppressed FI term, 
we take another different parameters.
Fig.~2 shows the case using $W_1$ with 
the parameters $p=2,b=1,d=-6,B=54,h=-\frac{1}{40}$ and
$\delta_{GS}=0.005$.
The dilaton is stabilized at $s = 7.1$, where we have 
$V_D \gg V_F$.
At this minimum, the total potential is obtained as
$V_T = 1.3 \times 10^{-11}$.
Now the potential value $V_T = 1.3 \times 10^{-11}$ 
is suppressed compared with $\delta_{GS}^2/s^3$.
In this case, the nonperturbative K\"ahler potential 
plays a role in suppressing the FI term.
Thus, the non-perturbative K\"ahler potential is useful 
to lead to a suppressed FI term when we take proper 
parameters.
The value of minimum potential depends linearly on 
$\delta_{GS}^2$.
When we take more suppressed value of $\delta_{GS}^2$, 
we obtain much smaller potential value.
\begin{figure}
\epsfxsize=0.7\textwidth
\centerline{\epsfbox{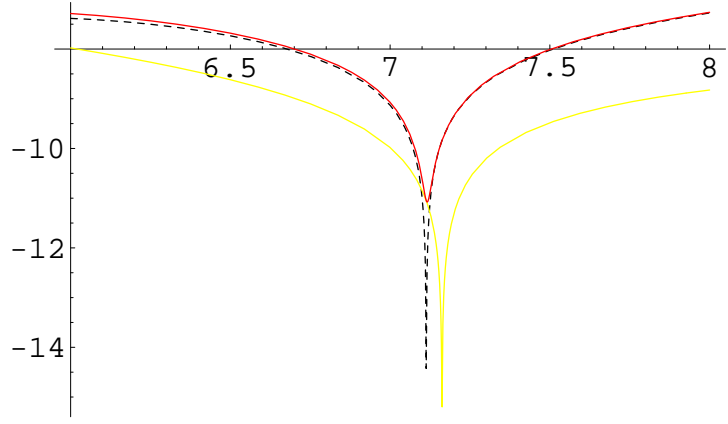}}
\caption{The potentials
for $W_1$ with parameters, $p=2,b=1,d=-6,B=54,h=-\frac{1}{40}$ and
$\delta_{GS}=0.005$.
A horizontal axis represents $s$ and a vertical axis represents
potentials.
When $s\simeq 7.1$, $V_T \simeq 1.3 \times 10^{-11}$.
(light - $V_D$, dotted - $V_F$,
bold - $V_T$)}
\end{figure}

Similarly, Fig.~3 shows the case using the superpotential
$W_2$ with the parameters, $p=3$,$b=1$, $d=4.5$, $\alpha =1$,
$\beta = 1$,
$b_1=15$, $b_2=12$ and
$\delta_{GS}=0.005$.
The dilaton is stabilized as $s=0.55$, where we have $V_D \gg V_F$.
At this minimum, the total potential is obtained as
$V_T = 2.8 \times 10^{-10}$, which is very suppressed
compared with $\delta_{GS}^2/s^3$.
The form of potential is steeper.
\begin{figure}
\epsfxsize=0.7\textwidth
\centerline{\epsfbox{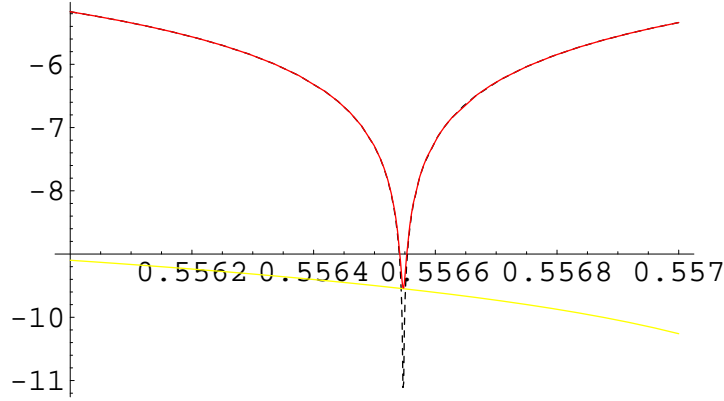}}
\caption{The potentials for $W_2$ with parameters,
$p=3,b=1,d=4.5,\alpha =1, \beta =3, b_1 =15, b_2 =12$ and $\delta_{GS}
=0.005$.
A horizontal axis represents $s$ and a vertical axis represents
potentials.
When $s \simeq 0.55$,
$V_T \simeq 2.8\times 10^{-10}$.
(light - $V_D$, dotted - $V_F$,
bold - $V_T$)}
\end{figure}

In the literature, another form of nonperturbative
K\"{a}hler potential has been used like
\begin{equation}
K = \ln ( e^{K_0} + e^{K_{np}} ).
\end{equation}
For this type, we can study similarly.
For example, Fig.~4 shows the case using this
K\"ahler potential and the superpotential $W_1$ with
the parameters, $p=1$, $b=1$, $d=-1$, $B=54$, $h =-1$ and
$\delta_{GS} = 0.02$.
The dilaton is stabilized at $s=6.0$,
where we have $V_D \gg V_F$.
At this minimum, the total potential $V_T = 1.1 \times 10^{-11}$
is very much suppressed compared with $\delta_{GS}^2/s^3$.
Therefore, it is possible to stabilize the dilaton
such that $\delta_{GS}^2/s^3 \gg V_D \gg V_F$ when we
take suitable values of parameters.
In this model, the potential value $V_T$ depends linearly on 
$\delta_{GS}^2$, again.
When we take smaller values of $\delta_{GS}$, we obtain 
more suppressed value of $V_T$.
\begin{figure}
\epsfxsize=0.7\textwidth
\centerline{\epsfbox{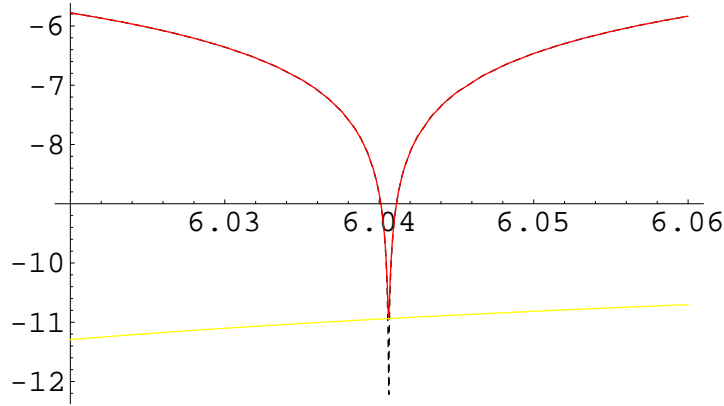}}
\caption{The potentials for $W_1$ with parameters,
$p=1, b=1, d=-1.5,B=54,
h=-1$ and $\delta_{GS}=0.02$.
A horizontal axis represents $s$ and a vertical axis represents
potentials.
When $s\simeq 6.0$, $V_T \simeq 1.1 \times 10^{-11}$.
(light - $V_D$, dotted - $V_F$,
bold - $V_T$)}
\end{figure}


We have neglected the perturbative K\"ahler term $K_p$.
Here we investigate its effects on our results.
We replace $K_0(S + \bar S)$ as follows,
\begin{equation}
K_0(S + \bar S) \rightarrow K_0(S + \bar S) + \frac{a}{S + \bar S},
\end{equation}
in the above models.
Then we repeat the above potential analysis.
The results, the stabilized values of $s$ and 
the potential values, are the same for $|a|< 0.1$ 
in all of the above models except the model shown in Fig.~3.
In the model of Fig.~3, the stabilized value of $s$ does not change,
but the potential value changes, that is, 
the total potential value $V_T$ at the minimum is 
obtained as $V_T \sim 10^{-7}$, e.g. for $a=0.01$.
This potential value $V_T \sim 10^{-7}$
is still suppressed compared with $\delta_{GS}^2/s^3$.
Thus, the nonperturbative K\"ahler potential 
contributes to suppress it.
However, the perturbative term in the K\"ahler potential 
is important in this model.
Its reason is that the stabilized value $s$ is small, while 
the dilaton is stabilized at larger values in the other models.
Hence, we have to take into account effects due to 
the perturbative K\"ahler term, in particular the models with 
a small value of $s$.
Furthermore, there is a possibility that  
the perturbative K\"ahler term plays a role to lead to 
suppressed FI term by cancelling the contribution due to 
$K_0$ in models with little contribution due to $K_{np}$.
In such case, the stabilized value of $s$ would be quite small like 
$s \sim a$.

Finally, we give a brief comment on the cosmic string problem.
After inflation, in the true vacuum the cosmic strings are
generated,{\footnote{Recently, the $D$-term inflation model
without generating cosmic strings has been
studied \cite{Urrestilla:2004eh}.}}
and that leads to the density perturbation, which is
inconsistent with the CMB observation for
$\lambda \sim 1$ if the FI term $\xi$ in
the true vacuum is obtained as $\sqrt\xi =O(10^{15}- 10^{16})$ GeV.
One way to avoid this problem is to take a suppressed value of
$\lambda$ like $\lambda = O(10^{-4} - 10^{-5})$
\cite{Endo:2003fr}.
In string theory, the coupling $\lambda$ also depends on
the dilaton and other moduli fields.
For example, in orbifold models, it is obtained as
$\lambda \sim e^{- a T}$, where $a$ is a constant
factor \cite{Hamidi:1986vh,Burwick:1990tu,Kobayashi:2003vi}.
That implies that a large value of $T$, e.g. $aT \sim 10$,
leads to a small value of $\lambda$.{\footnote{Furthermore,
the coupling $\lambda$ also depends on continuous Wilson lines
similarly \cite{Kobayashi:2003vi}.
Large value of continues Wilson lines lead to a suppressed value
of $\lambda$.}}
However, the inflaton must take a large expectation value of order $M_p$
in a model with such a extremely small $\lambda$. In this sence,
this option might not be a excellent solution.

The second possibility is that the FI term $\xi$ is suppressed
through the above mechanism
both in the false vacuum during inflation and in the true vacuum
like $\sqrt \xi < O(10^{15}- 10^{16})$ GeV.
In this case, the density fluctuation generated by cosmic strings
is suppressed sufficiently.
However, at the same time, the density fluctuation due to the
inflaton is also suppressed.
We need another source of the density fluctuation to be
consistent with the CMB observation, e.g. the curvaton.

We have another possibility, that is, the
dilaton is stabilized through the above mechanism
leading to $\sqrt \xi = O(10^{15}- 10^{16})$ GeV during inflation.
After inflation, the superpotential can change e.g.
the VEV of $\phi_-$ in the true vacuum can make
mass terms of hidden matter fields relevant to nonperturbative
superpotential.
In this case, the stabilized value of $s$ in the true vacuum would
be different from one in the false vacuum during inflation.
The true vacuum can correspond to more suppressed values of
$K_S$ and $\xi$ than those during inflation.
Analysis of such possibility in explicit models is
beyond our scope. It will be studied elsewhere.
Here we would like to mention that the nonperturbative K\"ahler
potential may solve not only the problem of the magnitude of the FI term
but also the problem of cosmic string in D-term inflation.

To summarize, we have studied the possibility for
suppressing the FI term in the $D$-term inflation scenario.
That is possible when we consider the nonperturbative
K\"ahler potential of the dilaton field.
That is an interesting possibility.
Thus, it would be important to study its implications
in detail by investigating evolutions of dilaton and other moduli
after inflation, a concrete model of gaugino condensation, and so on.

\section*{Acknowledgment}

One of the authors (T.K.) would like to thank Kiwoon Choi
for useful discussions.
T.~H.\/ and T.~K.\/ are supported in part by the Grant-in-Aid for
the 21st Century COE ``The Center for Diversity and
Universality in Physics'' from Ministry of Education, Science,
Sports and Culture of Japan.
T.~K. is supported in part by the Grant-in-Aid for
Scientific Research from Ministry of Education, Science,
Sports and Culture of Japan (\#14540256).
O.~S. is supported by National Science Council of Taiwan under
the grant No. NSC 92-2811-M-009 -180-.


\end{document}